
\input phyzzx
\def\co{cohomology\ }
\def\W{$W_N\ $ }
\def\G{${G\over G}\ $}
\def\GT{${G\over G}$ TQFT\ }

\def\c#1#2{\chi_{#1}^{#2}}
\def\p#1#2{\phi_{#1}^{#2}}
\def\b#1#2{\beta_{#1}^{#2}}
\def\g#1#2{\gamma_{#1}^{#2}}
\def\r#1#2{\rho_{#1}^{#2}}
\def\jt#1#2{{J^{(tot)}}_{#1}^{#2}}
\def\y{|phys>}
\def\t{\tilde}
 \def\pa{\partial}
\def \f{f^a_{bc}}
\def \fcr{\f \c  {} b \r  {} c}
\def \jg{J^{(gh)}}
\def \jb{J^{(BRST)}}
\def \dij{\delta_{i,j}}
\def \sij{\sum_{1\leq i\leq j\leq N-1}}
\def \pij{\prod_{1\leq i\leq j\leq N-1}}
\def \gij{g^{ij}}
\def\i{{(ij)}}

\def\S#1{$SL({#1},R)$}
\def\A {$A_{N-1}^{(1)}$}
\def\lj{\lambda_J}
\def\li{\lambda_I}
\def\cmp#1{{\it Comm. Math. Phys.} {\bf #1}}

\def\pl#1{{\it Phys. Lett.} {\bf #1B}}

\def\np#1{{\it Nucl. Phys.} {\bf B#1}}

\def\jmath#1{{\it J. Math. Phys.} {\bf #1}}

\REF\SY{M. Spigelglas and S. Yankielowicz
`` \G Topological Field Theories by
Coseting $G_k$";``Fusion Rules As Amplitudes in $G/G$ Theories,''}
\REF\Wgg{E. Witten, \cmp {144} (1992) 189.}
\REF\us  {O. Aharony, O. Ganor, N. Sochen,  J. Sonnenschein and S.
Yankielowicz ,`` Physical states in the \G models and two dimensional
gravity", TAUP- 1961-92 April 1992.}
\REF\BMP{P. Bowknegt, J. Mc Carthy and  K. Pilch Cern Preprint
TH-6162/91.}
\REF\Walgebra{See for instance
 C.N. Pope,
``A Review of W Strings"
 CTP-TAMU-30-92, Apr 1992, and references therein.}
\REF\FaLu{ Fateev and Lukeanov
{\it Int. Jour. of Mod. Phys. }{\bf A31} (1988)
507.}
 \REF\BeO{M. Bershadsky and H. Ooguri \cmp {126}  (1989) 49.}
\REF\GK{  K. Gawedzki and A. Kupianen , \pl {215} (1988) 119, \np
     {320} (1989)649.}
\REF\KS{D. Karabali and H. J. Schnitzer, \np {329} (1990) 649.}
\REF\BBSS{ See for instance F.A. Bais, P. Bouwknegt, M. Surridge and K.
Schoutens ``Extended Virasor Algebras" in Proceeding of the Niels Bohr
Institute Meeting
``Perspective in String Theory" Edited by P. Di Vecchia and
J. L. Petersen, and references therein.}
    \REF\DVV{R. Dijkgraaf, E. Verlinde, and H. Verlinde,
\np {352} (1991) 59;
``Notes On Topological String Theory And 2D Quantum Gravity,'' Princeton
preprint PUPT-1217 (1990).}
\REF\SoYa{J. Sonnenschein and S. Yankielowicz `` Novel Symmeetries in
Topological Conformal Field Theories" Tel Aviv Univ. preprint
TAUP-1898-91.}
 \REF\FeFr{B. Feigin and E. Frenkel \pl {246} (1990) 75.}
\REF\KaSU{Y. Kazama and H. Suzuki, \np {321} (1989) 232.}
\REF\WiKZ{E. Witten `` The N Matrix Model and Gauged WZW Models"
IASSNS-HEP-91/26.}
\REF\NeWa{D. Nemeschansky and N. P.
Warner ``Topological Matter, Integrable Models
 and Fusion Rings"  USC-91/031.}
\REF\BF  {Bernard   and G, Felder \cmp {}  (1991)  145.}
\REF\BMPN{P. Bouwknegt, J. McCarthy and  K. Pilch \pl {234} (1990) 297,
\cmp {131} (1990) 125.}
\REF\NaSu{T. Nakatsu and Y. Sugawara
``Topological gauged WZW models and 2 D gravity" Univ. of Tokyo preprint
UT-598 March 92,
BRST-Fixed points and Topolological Conformal Symmetry"
Univ. of Tokyo preprint
UT-599 March 92, }
\REF\NY{D. Nemeschansky and S. Yankielowicz
``N=2 W algebras, Kasama-Suzuki
Models and Drinfeld-Sokolov Reduction"
   USC-007-91.\break
K. Ito \pl {259} (1991) 73\break\hfill
 T. Inami , and  H.  Kanno `` N=2 Super W Algebras and generalized
 N=2 SuperKdV Hierarchies based on 2 Lie Superalgerbas"
 YITP-K-928, May 1991}
\REF\HuYu{ H. L. Hu and M. Yu ``On BRST cohomology of
SL(2)/SL(2) Gauged WZWN
models" Academia Sinica preprint AS-ITP-92-32.}

\REF\KK{V. G. Kac and  D. A. Kazhdan
{\it Adv. Math } {\bf  34}  (1979) 79.}

\rightline{TAUP- 1977-92}
\date{June 1992}
\titlepage
\vskip 1cm
\title{ \G Models and  \W strings}
\author {O. Aharony, J. Sonnenschein and S. Yankielowicz
\footnote{\dagger}
{Work supported in part by the US-Israel Binational Science
Foundation and the Israel Academy of Sciences.}}
\address{ School of Physics and Astronomy\break
Beverly and Raymond Sackler \break
Faculty of Exact Sciences\break
Tel Aviv University\break
Ramat Aviv, Tel-Aviv, 69978, Israel}
\abstract{
We derive the BRST cohomology of the
\G topological model for the case of \A .
It is shown that   at level
  $k={p\over q}-N$  the latter
  describes the $(p,q)$ $W_N$ minimal model coupled
to $W_N$ gravity (plus some extra ``topological sectors").}
\endpage

\overfullrule=0pt
Recently the space of physical states of the \G
topological models\refmark{\SY,\Wgg} for the case of $A_1^{(1)}$ was
extracted.\refmark\us\    By applying the method  of ref. [\BMP]   the
cohomology of the  gauge symmetry BRST operator  was
computed. Using a twisted
energy-momentum tensor  an intriguing correspondence
between the ${SL(2)\over SL(2)}$ model with level
$k={p\over q}-2$ and $(p,q)$
models coupled to gravity was established.
In the present work we extend this
analysis to  the \A  algebra and show that the model describes $W_N$
strings.\refmark\Walgebra\  More precisely,
the  $k={p\over q}-N$ level model
contains  $(p,q)$  $W_N$ minimal  models\refmark\FaLu
 coupled to $W_N$  gravity with some
extra ``topological sectors".
We shall present some evidence which support this result and also work out
explicitely the BRST cohomology. Related results were obtained previousely
within the Hamiltonian reduction approach,
\refmark\BeO which has common features
to our appraoch. In particular it has been
shown that the \S N Kac-Moody algebra
at level $k={p\over q}-N$ gave rise to the  $(p,q)$  $W_N$
algebra.\refmark\BeO

  The  \GT is constructed by gauging the anomaly free
diagonal $G$ group of the WZW model.
 The quantum action  of the  model was
 shown to be composed of three decoupled
parts:\refmark{\GK,\KS}
  $S_k(g)-$ a $WZW$ model of level $k$
  with $g\in G$, $S_{-(k+2C_G)}(h)-$ a
$WZW$ model of level
$-(k+2C_G)$ with $h\in G$,
 and a  dimension $(1,0)$ system of anticommuting ghosts $\rho$ and $\chi$
  in the adjoint representation of the group. The action, thus, reads
 $$S_k(g,h,\rho,\chi) =S_k(g) +S_{-(k+2C_G)}(h) -i\int d^2z
Tr[\bar\rho \pa \chi + \rho\bar\pa\chi], \eqn\mishwzwh$$
 where  $C_G$ is the second Casimir of the
adjoint representation. For fractional
level we take for $G$ the non-compact
group \S N.
In this case parametrizing   the gauge fields by orbits of $G_C$, the
complexificaiton of $G$, is not
permissible\footnote{*}{ We thank E. Witten for
paying our
attention to this point.} and thus the  naive passage from the
gauged WZW action to the one given in eqn. \mishwzwh\ may  not hold.
Nevertheless,  one can always take  the action of eqn. (1) as the
definition of the model.

Invariance of each of the three terms under holomorphic
$G$ transformations gives rise to  three  Kac-Moody currents
$J(z)= g^{-1}\pa g$, $I(z)= h^{-1}\pa h$ and ${\jg}^a=\fcr$
of levels  $k$,$-(k+2c_G)$ and $2c_G$
respectively.

We decompose the $J^a$ currents ( and similarly the  $I^a$ and
${\jg}^a$) to those  related to the Cartan sub-algebra generators
and to those related to the positive and negative roots.
In the present work we are interested
in \A and thus there are $N-1$ of the former which we denote by $J^i$,
$i=1,...,N-1$ and ${N(N-1)\over 2}$ of the latter which we  denote by
$J^{\pm(ij)}$. Each of the $J^{\pm(ij)}$ corresponds to a root $\pm(
\alpha_i+\alpha_{i+1}+...+\alpha_{j-1}+\alpha_j)$
where the simple roots are
$\alpha_1,...,\alpha_{N-1}$ and with $1\leq i\leq j\leq N-1$.
 The operator product algebra takes now
the form
$$\eqalign {J^i(z)J^j(\omega) =& {\gij k\over
(z-\omega)^2}+O(z-\omega)\cr
 J^{i}(z)J^{\pm (jk)}(\omega) =&
 {f^{i\pm(jk)}_{\pm(lm)}J^{\pm(lm)}(\omega)\over
(z-\omega)}  +O(z-\omega)\cr
J^{+\i}(z)J^{-(kl)}(\omega) =&  \delta^{ik}\delta^{jl}[{k\over
(z-\omega)^2}+
{f^{+(ij)-(kl)}_mJ^m(\omega)\over
(z-\omega)}]+{f^{+(ij)-(kl)}_{\pm(mn)}J^{\pm(mn)}(\omega)\over
(z-\omega)}+ O(z-\omega),\cr
J^{\pm\i}(z)J^{\pm(kl)}(\omega) =&
{f^{\pm(ij)\pm(kl)}_{\pm(mn)}J^{\pm(mn}(\omega)\over
(z-\omega)}+O(z-\omega),\cr
}\eqn\mishJJ$$
where   $\gij$ is the inverse Cartan matrix,
given by  $\gij={1\over
N}Min(i,j)(N-Max(i,j))$, and $f^{ab}_c$ are the
group structure constants.

We define now $\jt{} {a} $
$$\jt {} {a}   =J^a +I^a +{\jg}^a= J^a +I^a +i\fcr \eqn\mishbJ$$
which obeys a Kac-Moody algebra of level
$$ k^{(tot)} = k  -(k+2c_G) + 2c_G =0. \eqn\mishbk$$
The total  energy-momentum  tensor  $T(z)$ is a sum of  Sugawara terms
of the $J^a$ and  $I^a$ currents and the usual contribution of a
$(1,0)$  ghost system, namely\refmark{\GK,\KS}
$$T(z) = {1\over k+c_G }:J_a J^a: -
{1\over k+c_G }:I_a I^a: +\r  {a} {} \pa \c   {} a. \eqn\mishbT$$
The corresponding  Virasoro central  charge  vanishes
$$ c^{(tot)} = {k d_G\over k+c_G}  -{(k+2c_G) d_G\over -(k+2c_G)+c_G}
-2d_G =0 \eqn\mishbk$$
as it is the case for  any TCFT. In each of the three sectors of the model
one can generalize the symmetry generator $T(z)\equiv W^{(2)}(z)$  to
$W^{(l)}$ for $l=2,...,N$. $W^{(l)}$
is built as  a symmetric normal ordered
$l^{th}$ product of the currents
associated with the $l^{th}$ Casimir operator.
 For instance  the purely $J$ sector contribution to $
W^{(3)}$\refmark\BBSS
 will be
$$W_J^{(3)} =\eta^{(3)} d^{(abc)}: J^aJ^bJ^c : \eqn\mishWW$$  where
$d^{(abc)}$ is the symmetric traceless
tensor invariant of the underlying Lie
algebra, and $\eta^{(3)}=\sqrt{N}[3(N+k)\sqrt{2(N+k)(N^2-4)}]^{-1}$. The
generator $W^{(3)}$ can then be constructed from $W_J^{(3)},$
$W_I^{(3)}$,
$W_{\jg {} {}}^{(3)} $ and similar cubic  terms with mixed
currents along the lines of ref.[\BBSS].

  It is easy to realize that the basic  algebraic
structure of a TCFT\refmark{\DVV,\SoYa} is obeyed by the model.
This is expressed in terms of two bosonic and two fermionic operators.
The former are  $T(z)$ and the ``ghost number current" $J^\#= \c a {}
\r {} a $. The
fermionic currents are a
dimension one current which is the BRST current $\jb$
and a dimension two operator $G$. These  holomorphic symmetry  generators
are given by
$$ \eqalign{ \jb=&\c a {} [J^a + I^a + \half {\jg}^a]  \cr
G=&{1\over k +c_G}\r a {} [J^a - I^a ]  \cr}\eqn\mishbGJ$$
The TCFT algebraic structure now reads
 $$\eqalign{ T(z) =&\{ Q, G(z)\} \cr  \jb =&\{ Q ,
j^\#(z)\}\cr}
\eqn\mishbalgebra$$
where $Q=\oint \jb(z)$ is the BRST charge. In addition to $T(z)$ and
$\jb(z)$, the total current $\jt {} a $ is also BRST exact,
$$ \jt {} a (z) = \{ Q , \r {} a  \}.  \eqn\mishbjt $$
An essential step for the extraction of the physical states\refmark\us
of the theory and for the
comparison with $(p,q)$ $W_N$ string models, is the
parametrization   of the currents in terms of free fields.
For the $J^a$ sector   we introduce a set of scalars $\phi_i, \
i=1,...,N-1$ with $\phi_i(z)\phi_j(\omega) =-\dij log (z-\omega)$
 and a commuting $(1,0)$ system
 $\b {} \i , \g {} \i $ $(i\leq j)$ for each
positive root obeying
$\g {} \i (z)\b {} {(kl)} (\omega) =\delta^{ik}\delta^{jl} {1\over
(z-\omega)}$.
The explicit form of  the $J_n^i$ and
$J_n^{+\i}$ currents in terms of these free
fields\refmark{\FeFr,\BMPN} is given by
$$ \eqalign{ J_n^{+\i}=&\b n \i
-\sum_m\sum_{k=1}^{i-1}:\g m {(k,i-1)}\b {n-m}
{(kj)}:\cr
J_n^i =& \sum_{j=1}^{N-1}\gij [\nu\vec\alpha_j\cdot\vec \p n {} +
2\sum_m :\g m {(jj)} \b {n-m} {(jj)}:\cr -&\sum_m\sum_{k=1}^{j-1}(:\g
m {(k,j-1)}\b {n-m} {(k,j-1)}:-:\g m {(kj)} \b {n-m} {(kj)}:)\cr
 -&\sum_m\sum_{k=j+1}^{N-1}(:\g
m {(j+1,k)}\b {n-m} {(j+1,k)}:-:\g m {(jk)} \b {n-m} {(jk)}:)]\cr}
\eqn\mishJij$$
These expressions as well as those for the negative root currents
are derived
 by applying  the commutation relations for
  the currents which
 correspond to the
simple roots:
$$ \eqalign{ J_n^{+(ii)}=&\b n {(ii)}
-\sum_m\sum_{k=1}^{i-1}:\g m {(k,i-1)}\b
{n-m} {(ki)}:\cr
 J_n^{-(ii)} =& -\sum_m\g m {(ii)} \nu\vec\alpha_i\cdot\vec \p {n-m} {} +
(k+i-1)n\g n {(ii)}\cr
 + &\sum_m\sum_{k=i+1}^{N-1}:\g
m {(i,k)}\b {n-m} {(i+1,k)}:
 -\sum_m\sum_{k=1}^{i-1}:\g
m {(k,i)}\b {n-m} {(k,i-1)}:\cr
-&\sum_{l,m}:\g {n-m-l} {(ii)}[\sum_{k=i}^{N-1} \g {m}
{(ik)}\b l {(ik)}- \sum_{k=i+1}^{N-1} \g {m}
{(i+1,k)}\b l {(i+1,k)}]:\cr}\eqn\mishJii$$
with $\nu^2 =k+N$.
The free field parametrization of the $I$ sector
is done using a similar set of
fields denoted by $\t\phi, \t\b {} {} $ and $\t\g  {} {}$. Our
experience with the $A_1^{(1)}$ case taught us that in order to follow the
analysis of ref. [\BMP] we would have to perform an involution, namely,
$I_n^i\leftrightarrow  -I_n^i\ , I_n^{+\i}\leftrightarrow  I_n^{-\i} $,
 $  k\rightarrow -k-2N $,  $\nu\rightarrow i\nu$
and then invoke the free field parametrization. Alternatively one can
perform the involution in the $J$ sector and use
for  the $I$ sector the analogs
of eqns. \mishJii,\mishJij.

The correspondence between $SL(2,R)$ \G models of level $k={p\over q}
-2$ and $(p,q)$ minimal models coupled to
gravity was demonstrated\refmark \us
using  a twisted energy-momentum tensor. The generalization
of the
latter to  the \S N case  is given by
$$T(z)\rightarrow \t T(z) =
T(z) +\sum_{i=1}^{N-1}\pa\jt {} i (z)\eqn\mishetT$$
Obviously since $T(z)$ and $\pa\jt {} i (z)$
are BRST exact so is $\t T(z)$.
Using eqn. \mishJJ\ it is easy to verify that the
Virasoro central charge of
each sector is shifted  to
$$c\rightarrow \t c = c -12k\sum_{i,j} g^{ij}= c-d_Gc_Gk =
c-(N-1)N(N+1)k,\eqn\mishtc$$
where $k$ is the level of that sector.

The twisted  ghost system will be  shown to include the ghosts of a $W_N$
gravity, namely, a sequence of ghosts with dimensions
$(i,1-i)$ for $i=2,...,N$
contributing $\t c_{Wgh}=-2(N-1)[(N+1)^2+N^2] $ to $\t c$.
The rest of the
ghosts are paired with commuting fields of  the same conformal structure
coming from the  $J$ and $I$ sectors. Therefore, the net matter degrees of
freedom have the following  Virasoro anomaly  $c=\t c_J-\half[\t
c_{(gh)}-\t c_{Wgh}]=(N-1)[(2N^2 +2N +1) -N(N+1)(t+{1\over
t})]$ which is exactly that of a $(p,q)$ minimal $W_N$ matter
sector\refmark\FaLu provided  $t\equiv k+N={p\over q}$.
This can be explicitly verified by analyzing the dimensions
and contributions
to $\t c$ of the various free fields in the $J$ sector.
Recall that before
twisting $(\b {} {(ij)},\g {} {(ij)})$ are commuting $(1,0)$
systems and the
contribution of the set of all  $\p{} i$ to $c$ is $c_\phi = (N-1)
-{12\sum_{i,j}g^{ij}\over k+N}= -(N-1)[{(N^2+N)\over t} -1]$.
Due to the twisting the  $(\b {} {(ij)},\g {} {(ij)})$ systems acquire
dimensions of $(i-j, j-i+1)$ and thus there are $N-1 $  systems of dimension
$(0,1)$,   $N-2$ pairs of fields of dimension  $(-1,2)$
up to one pair of
dimensions $(2-N,N-1)$ where we have used  the bosonization of  eqns.
\mishJii,\mishJij\  in this sector.
The $\phi$ central charge is modified to
$$\t c_\phi
=(N-1)[(2N^2 +2N +1) -N(N+1)(t+{1\over t})]\eqn\mishcm$$ This result  is
identical to the net matter contribution to $c$ given above and hence
the $\phi^i$ fields are in fact those of the $W_N$ model. A further
indication of this        equivalence  is
 the dimensions of the $\phi$ fields which correspond to the maximal
weights $\lambda^j=\sum_k g^{jk}[(r_k-1)-t(s_k-1)]$ where $\lambda^j$
are defined below in eqn. (20).       After the twisting the dimensions
are
$$\Delta_{r_1,...,r_{N-1},s_1,...,s_{N-1}} = { 12\sum_{i,j} g^{ij}
(ps_i-qr_i)(ps_j-qr_j)-N(N^2-1)(p-q)^2\over 24 pq}\eqn\mishDelta$$
as in the
$W_N$ minimal models.\refmark\FaLu\  If one parametrizes the $I$
sector in the
same way as  the $J$ sector, then  clearly  the modified dimensions of the
$(\t\b {} {(ij)},\t\g {} {(ij)})$ fields are the same as
those without tilde.
{}From the point of view of their contribution to $\t c$   the $\t \p {} i$
fields  are then identical to those of $W_N$ gravity. This is achieved by
replacing $t$ with $-t$ in $\t c_\phi$ defined above.  Using an inverse
parametrization in this sector,  the $W_N$ gravity modes
and the fields which
pair with the redundant ghosts are not any
more the $\t \phi^i$ and $(\t\b {}
{(ij)},\t\g {} {(ij)})$ fields respectively
but rather some combination  of
them.
 With respect to the untwisted $T$ the ghosts $(\r {} a,\c {} a)$ are
all of dimension $(1,0)$. The   ghost part of $\t T(z)$ has the form
$$ \eqalign{\t T^{(gh)}(z) =&g_{ab}:\r {} a \pa \c {} b:\cr
+&\sum_{1\leq i\leq j\leq N-1} (j-i+1)\pa[:\r {} {-(ij)}(z)
\c {} {+(ij)}(z):-
:\r {} {+(ij)}(z) \c {} {-(ij)}(z):].\cr}\eqn\mishTgh$$
It is thus obvious that
the members of the Cartan sub-algebra $\r {} i, \c {} i$
remain $(1,0)$ fields.
On the other hand the pair $(\c {} {+(ij)}, \r {} {-(ij)})$
carries now
dimensions $(i-j-1, 2+j-i)$ and  $(\c {} {-(ij)}, \r {} {+(ij)})$ carry
dimensions $(j-i+1, i-j)$. Altogether one finds for the
$(\chi, \rho)$ ghosts
$N-1$ pairs of fields of dimension $(0,1)$ coming from
the Cartan sub-algebra,
$N-1$ pairs of dimension $(-1,2)$,
$N-2$ pairs of dimension $(-2, 3)$ up to a  pair  of  dimension $(1-N,N)$, and
similarly $N-1$ pairs of dimension $(1,0)$ up to one pair of dimension
$(N-1,2-N)$.
It is now clear that  when the dust settles the \G model
of  \S N at level
$k={p\over q}- N$ has the field content of a minimal  $W_N$ $(p,q)$ model
coupled to  $W_N$ gravity  plus pairs
of ``topological sectors" namely pairs of
commuting and anti-commuting $(i,1-i)$ ghost
systems for $i=1,...,N-1$.

Next we  proceed to extract  the
space of  physical states of the model.  The
physical states   are in  the  \co of  $Q$, the BRST  charge , $|phys>\in
H^*(Q)$.
 Since  both  $\jt n a $ and $L_n$
 are  $Q$ exact (eqn.\mishbalgebra,\mishbjt\
), it is clear that
$$L_0\y =  0 \qquad \jt 0 i \y =  0 \  \ \ \  (i=1,...,N-1)\eqn\mishL.$$
and therefore we can work on the space annihilated by $L_0$ and $\jt 0 i$.
Note that since $\jt 0 i$ is $Q$ exact
the physical states before and after
twisting are the same. We can, therefore, concentrate on the untwisted \G
theory.   The BRST charge can be decomposed
into  $$\eqalign{Q  &= \c 0 i \jt 0
i  + M^i \r 0  i +\hat Q \cr M^i &=
-\half f^i_{bc}\sum_{n}  :\c{-n} b\c{n}c :
\cr} \eqn\mishhatQ$$
where the $i-$ sum  is over the Cartan subalgebra,
so that on the sub-space  of states annihilated by   $\r
0 i$,  $Q=\hat Q$.
  We  build the states of  $H^*(\hat Q)$
  on a highest weight vacuum  $|J,I>$
obeying the following relations
 $$\eqalign{J_{n>0}^a|\vec\lj,\vec\li>=&0
 \qquad J_0^{+(ij)}|\vec\lj,\vec\li>=0
\qquad J_0^i|\vec\lj,\vec\li>=\lj^i|\vec\lj,\vec\li>\cr
I_{n>0}^a|\vec\lj,\vec\li>=&0\qquad I_0^{+(ij)}|\vec\lj,\vec\li>=0 \qquad
I_0^i|\vec\lj,\vec\li>=\li^i|\vec\lj,\vec\li>\cr
\c{n>0} a|\vec\lj,\vec\li>=&0\qquad\r{n\geq 0}
a|\vec\lj,\vec\li>=0.\cr}\eqn\mishIJ$$ The  weights in this
parametrization are $\sum \vec\alpha_i \lj^i$ and
$\sum \vec\alpha_i \li^i$,
respectively. Using  the free field parametrization  of the
currents \mishJij,  it is straightforward to check that $\b {0}
{\i}|\vec\lj,\vec\li>= \t\g {0} {\i}|\vec\lj,\vec\li>=0$.
In order  to apply the spectral sequence
decomposition\refmark\BMP of the BRST
charge, we have to assign a degree  to the various
fields.
We use an assignment that generalizes the one used for the $A_1^{(1)}$
case\refmark\us
and which produces as the lowest degree term in $Q$ a term with
zero  degree.

This assignment
reads $$\eqalign{ &deg (\c {} {a}) =deg (\g  {}{ \i}) =deg
(\t\g {} { \i}) =deg (\p {} {i+}) =1\cr
 & deg (\r {} {a})= deg (\b {} { \i}) =deg (\t\b {}{ \i}) ) =deg (\p {}
 {i-}) = -1\cr}\eqn\mishdeg$$
where   $\p n {i\pm} ={1\over \sqrt{2}}(\p n {i}  \pm i\t\p n {i})$.
The  decomposition of $\hat Q$ into
terms of different degrees is given by
$$\eqalign {\hat Q=& Q^{(0)} +Q^{(1)}+... +Q^{(N+1)} \cr
Q^{(0)} =&\sum_n \sij(\c {-n} {-\i}  \b n \i +
\c {-n} {+\i} \t\b n \i )
+\sqrt{2}\nu
\sum_{n\ne0}\sum_{i=1}^{N-1} \c {-n} i(\vec\alpha_i\cdot  \vec\p n -
)\cr}\eqn\mishdegQ  $$
where
 $deg(Q^{(n)})=n$. We do not write down  expresions for
$Q^{(n\neq 0)}$ since they would not be needed for the
extraction of physical states.  Let us now examine  the cohomology
$H^*(Q^{(0)})$.
The conformal dimension operator $L_0$ when acting on a given
state has two parts: one that determines the level of the vacuum state and
the other that  determines the dimension  of the excitations
$$\eqalign{\L_0 =
&\hat L_0 +{1\over 2(k+N)}
[\sum_{i,j}g_{ij}(\lj^i\lj^j-\li^i\li^j)-2\sum_i(\lj^i-\li^i)]\cr
 \hat L_0=&\sum_n\sij n [:\b {-n} \i \g n \i: +:\t\b {-n} \i\t\g n \i:+
:\r {-n} {+\i} \c n {-\i}: +:\r {-n} {-\i}\c n {+\i}:]\cr
 +&\sum_{n\neq 0}\sum_{i=1}^{N-1}
 [n\sum_{j=1}^{N-1}g_{ij}:\r {-n} i\c n  j:  +
:\p {-n} {i+} \p n {i-} : ]\cr} \eqn\mishlzero$$
In analogy to the $SL(2)$ \G model\refmark\us one can easily verify that
 $\hat L_0$, the contribution to $L_0$ of
the
excitations,  is $Q^{(0)}$ exact
$$ \eqalign{\hat L_0 &= \{ Q^{(0)}, \hat G^{(0)}\}\cr
\hat G_0 &=-\sum_n\sij n[\r n {+\i} \g {-n} \i + \r n {-\i} \t\g {-n} \i]
 +{1\over \sqrt{2}\nu}\sum_{n\neq 0}\sum_{i,j,k=1}^{N-1} g_{ij}\r n i
\epsilon^k_j\p {-n} {k+}.\cr}\eqn\mishatG$$
where $ \epsilon_k^j$ are a set of
numbers obeying $ \sum_{i=1}^{N-1}\alpha_i^j\epsilon^k_i=\delta^{jk}$.
Hence, $\hat L_0$ annihilates the states
in the \co of  $Q^{(0)}$ on  the Fock space
and there are no excitations in
$H^*(Q^{0})$.
Since both $L_0$ and $\hat L_0$ annihilate
physical states so does $L_0-\hat
L_0$  and  therefore there is a restriction on the vacuum
$|\vec\lj,\vec\li>.$
The states built on this vacuum  may contain only the zero modes
$\t\b 0 \i ,\g 0 \i ,\c 0 {-\i} ,\c 0 {+\i} $.
The computation of ${Ker(Q^{(0)})\over Image (Q^{(0)})}$
for the $A_1^{(1)}$
is written down in ref. [\us]. An analogous computation for   \A  gives
$$ H^{rel} (Q^{(0)}) = \{\pij \c 0 {+\i} |\vec\lj,\vec\li> ;\   \ \
\lj^i+\li^i= -i(N-i)\} . \eqn\mishrelco$$
The condition $\lj^i+\li^i= -i(N-i)$  is
    analogous to
$I=-(J+1)$
 for the $SL(2)$ case.
The absolute cohomology (without the restriction $\r 0 i=0$ )is
$$H^{abs} (Q^{(0)}) \simeq H^{rel} (Q^{(0)})
\oplus\sum_{\{k_1,...,k_l\}}\c 0 {k_1}... \c 0 {k_l} H^{rel} (Q^{(0)})
\eqn\mishabsco$$ where the sum  is over $\{k_1,...,k_{l}\}$ which are  all
possible subsets of the set  $1,...,N-1$. Thus, each state in the relative
cohomology gives rise to $2^{N-1}$
states in the absolute cohomology.  Next we
want  to examine  whether
 $ H^* (Q^{(0)})\simeq H^*(Q)$.
 In ref. [\BMP] it was shown that the latter
holds when the degree is bounded on the Fock space built on
$|\vec\lj,\vec\li>$.
Since the physical states are annihilated by  $ L_0$, it is clear that
for a given $\vec\lj,\vec\li$ the
degree carried by the excitations is bounded. Hence, one has to   consider
possible contributions to the
total degree from zero modes. Obviously, there
can be only one zero mode for each $\c {} {}$ that corresponds to a root.
If we denote by $N_\i$ and $\t N_\i$ the number of $\g {0} \i$ and $\t\b
{0} \i$ respectively, the condition $\sum_i J_0^i =0$ bounds
$\sij (j-i+1)(N_\i +\t N_\i) $ from both sides. Since $N_\i$ and $\t N_\i$ are
non-negative it
implies that all of them are bounded separately and thus also their
contribution to the total degree
which is $\sij(N_\i -\t N_\i) $. We conclude
that using the lemmas of ref. [\BMP] the isomorphism $ H^* (Q^{(0)})\simeq
H^*(Q)$
holds, and thus in the cohomology
of $Q$ there is a single state built on the
vacuum  $|\vec\lj,\vec\li>$
 and it carries  ghost number $G={N(N-1)\over 2}$.

To  derive   the space of physical
states one has to  translate the  cohomology
on the Fock spaces into the space of
irreducible representations of \A in the
matter sector.
In analogy to the totally reducible representations of \S 2
\refmark\us,
there are similar ones for \S N provided $k+N = {p\over q}$   and
for particular
highest weights. A brief discussion of the latter is presented in
the appendix.

A generalization of the procedure of ref. [\BF]
enables   us to write down the
irreducible representations as the cohomology of a BRST like
operator $Q_J$,\refmark\BMPN acting on
a union of the  Fock spaces, at zero
degree.  The relevant cohomology  with ghost number  $G=n$
 is given by\refmark\us
$$H^{(n)}_{rel} [ H^{(0)}({\cal F}_{\vec\lj},Q_J)\times {\cal
F}_{\vec\li}\times {\cal F}^G,Q]\eqn\mishHa$$
where ${\cal F}_{\vec\lj}$
is the relevant union of Fock spaces and where  we
shift the definition of the ghost number by ${N(N-1)\over 2}$.
Due to the fact that $Q_J$ commutes with all the currents,
and that both cohomologies
are non-zero at a single degree, this cohomology  is
isomorphic to\refmark\us
$$H^{(n)} [ H^{(0)}_{rel}({\cal F}_{\vec\lj}\times
{\cal
F}_{\vec\li}\times {\cal F}^G,Q),Q_J].\eqn\mishHb$$
The cohomology in the brackets is exactly the one computed above. Hence,
for a given $\vec\li$, there is a  physical
state iff it is built on a vacuum $\t\lj$ contained in
${\cal F}_{\vec\lj}$ obeying $\t\lj^i+\li^i= -i(N-i)$ and
carrying a ghost number
equal to the degree of  $\t\lj$ in the $Q_J$ complex.
Denoting  $J_i= {1\over 2}\sum_{j=1}^{N-1} g_{ij} \li^j$, the condition on the
highest  weight state reads $$2{(J_i)}_{r_i,s_i} +1
=r_i-(s_i-1)(k+N)\eqn\mishJrs$$  where $\sum_{i=1}^{N-1}r_i<p,\
\sum_{i=1}^{N-1}s_i<q+N-1$.
In the $A_1^{(1)}$  model it was found\refmark\us
that there is a single state at each ghost number $G=-2l$ built on
$|J_{r,s},-J_{r+2lp,s}-1 >$ and a single state for  $G=1-2l$  built on
 $|J_{r,s},-J_{-r+2lp,s}-1 >$. In the present case of   \A  there is an
$N-2$ dimensional lattice of states   for each ghost number and $J$. This
follows
from the $N-1$ dimensional
lattice of Fock spaces which are derived by Weyl
reflections as well as shifts by
linear combination of  roots.\refmark\BMPN

In the present work
the space of physical states of the \G model based on \A was derived by
computing the cohomology of the gauge BRST charge on the space of  Kac-Moody
irreducible representations  in the ``matter sector". We showed that after
twisting, apart from some "topological sectors", the conformal properties of
the
fields of the model at level $k+N={p\over q}$ are the same as those of
$(p,q)$ $W_N$ strings. In the $A_1^{(1)}$ case\refmark\us with $k+2={p\over q}$
the partition function on the torus was shown  to be identical to that of the
$(p,q)$ minimal model coupled to gravity. In the present case one can repeat
this calculation by  inserting  the values of the $\hat L_0$ and $ \widehat\jt
0
i$
of the physical states into an index of the form $ Tr[(-)^Gq^{\hat
L_0}e^{i\pi\theta^i \widehat \jt 0 i} ]$, where $\theta^i$ are associated
with the moduli of
flat gauge connection on the torus.
For a particular value of the moduli this reduces  to
 $ Tr[(-)^Gq^{\t L_0}]$, where $\t L_0$ is the twisted one given in eqn.
\mishetT, and should reproduce
  the torus  partition function of the $(p,q)$ $W_N$ strings.
 An interesting question   is the relation
between the \A\  \G models and the twisted Kazama-Suzuki models.\refmark\KaSU
The explicit formulation of the Kazama-Suzuki models as gauged WZW models was
discussed in ref. [\WiKZ]. Moreover,   the  \G models  can be embedded into
topological  matter theories obtained by twisting hermitian symmetric  N=2
supersymmetric coset  models.\refmark\NeWa The Kazama Suzuki models based on
the
coset  ${SU(N)\over SU(N-1)\times U(1)}$ fall into this class of models.
In addition, it has been shown that these  models have $W_N$ algebra as their
chiral algebra.\refmark\NY It is interesting to
note that one can consider a whole series
of $N=2$ supersymmetric coset models by taking for $G=SU(N)$ the following
subgroups $H=U(1)^{N-1}, SU(2)
\times U(1)^{N-2},...,SU(N-1)\times U(1)$ with
fermions residing in the coset.
Upon twisting one obtains a corresponding series
of topological models. Formally the \G model will be the next one in this
series.  The cohomologies associated
with these models  and their relations to
the cohomology of the \G model is
under current investigation. Some related work
for the case of $H= U(1)^{N-1}$ can be found in ref. [\NaSu].
  Another issue
that has to be resolved is the
precise relation between the cohomology of the
Kac-Moody BRST operator and that of the Virasoro and $W_N$  cohomologies.
Resolving that question
may open the way to the use of the Kac-Moody BRST charge
( linear in the currents)  to deduce $W_N$ cohomologies.

Upon completion of this
work we  have received a work\refmark\HuYu  where the cohomology of the
$A_1^{(1)}$ \G case is discussed.
 \ack{ We would like to thank T. Eguchi for a useful conversation.
One of us J.S would like to thank W. Lerche and P. Bouwknegt for fruitful
discussions.}

\appendix
\centerline{Totally irreducible representations of \S N}

According to the Kac-Kazhdan
formula\refmark\KK a highest weight representation
$\vec\Lambda$ has a singular
vector of degree $n\vec\alpha$ for $n$  a positive
integer and $\vec\alpha$ a positive root of level $k$ \S N, iff
$$\Phi_{\vec\alpha,n} (\vec\Lambda)\equiv (\vec\Lambda+\vec \mu,
\vec\alpha)-{n\over 2}(\vec\alpha,\vec\alpha) =0\eqn\mishSin$$
where $\vec \mu$ is the sum of the fundamental  weights
which obey $(\Lambda_i,\alpha_j)=\delta_{ij}$
with $i,j=0,...,N-1$. Let us now
parametrize the highest weight representations as follows
$$\vec\Lambda =(k-2J_1-2J_2-...-2J_{N-1})\vec\Lambda_0
+\sum_{i=1}^{N-1}2J_i\vec\Lambda_i \eqn\mishLam$$

The positive roots of the \A algebra are
$$\eqalign{ \ \ (i)\ \alpha=&n'(\alpha_0+\alpha_1+...+\alpha_{N-1}) +
(\alpha_i+\alpha_{i+1}+...+\alpha_{j})
\qquad n'\geq0,\  1\leq i\leq j\leq N-1 \cr
(ii)\   \alpha=&n'(\alpha_0+\alpha_1+...+\alpha_{N-1}) \qquad n'\geq 1 \cr
(iii)\  \alpha=&n'(\alpha_0+\alpha_1+...+\alpha_{N-1}) -
(\alpha_i+\alpha_{i+1}+...+\alpha_{j})
\qquad n'\geq1,\  1\leq i\leq j\leq N-1 \cr}\eqn\mishalpha$$
 The simple roots( for  $N>2$) obey the scalar product

$$
 (\alpha_i,\alpha_j)=\cases{ 2& $i=j$\cr
          -1&           $i-j=\pm1\ ( mod\ N)$\cr
 0&             otherwise\cr}\eqn\mishscalar$$
Substituting eqn.\mishLam\ and eqn.\mishalpha\ for $\vec\Lambda$ and
$\vec\alpha$ respectively, the condition for
singular vectors eqn.\mishSin\
takes the form
$$\eqalign {\ \ (i) \Phi_{\vec\alpha,n}(\vec\Lambda) &=(k+N)n' - n + (
2J_i+2J_{i+1}+...+2J_j) +(j-i+1) =0\cr
\ (ii) \Phi_{\vec\alpha,n}(\vec\Lambda) &=(k+N)n' =0\cr
 (iii) \Phi_{\vec\alpha,n}(\vec\Lambda) &=(k+N)n' - n - (
2J_i+2J_{i+1}+...+2J_j) -(j-i+1) =0 \cr}\eqn\mishSinn$$
Equation (ii) has obviously a solution only for $k=-N$. Let us define  the
totally
reducible representation as the one that corresponds to the maximal
number of $n,n'$ which solve
equations (i) and (iii). Infinitely many solutions
for $n,n'$ exist provided $k+N={p\over q} $
where $p,q$ are two integers with
no common divisor. In that case for every
$n,n'$ solution, $n+p,n'+q$ also
solves the equations. In fact  $n+lp,n'+lq$ solves equation (i) and
$-(n+l'p),-(n'+l'q)$ solves (iii) when the
shifted numbers are still in the
allowed domains.
The  maximal number of solutions occurs therefore
only for $0\leq n'< q$ and $0 <n< p$.
 Denoting $n,n'$ for $i=j$ by $r_i,s_i$ one gets
$$ 2J_i +1 = r_i- (k+N)(s_i-1)\eqn\mishrs$$ which has the same form as the
condition for the singular state for the $A^{(1)}_1$  case.
Imposing the previous constraints
for all $\vec\alpha$ one finds that $r_i$
$s_i$ are positive integers that obey
$$\sum_{i=1}^{N-1} (s_i-1) < q \  \  \ \ \  \sum_{i=1}^{N-1} r_i < p
\eqn\mishsr$$.
  \refout\end
\bye